\documentclass[english]{IEEEtran}
\usepackage[T1]{fontenc}
\usepackage[latin9]{inputenc}
\usepackage{amsthm}
\usepackage{amssymb}
\usepackage{esint}

\makeatletter
\theoremstyle{plain}
\newtheorem{thm}{\protect\theoremname}
\theoremstyle{remark}
\newtheorem{rem}[thm]{\protect\remarkname}

\usepackage{cite}

\makeatother

\usepackage{babel}
\providecommand{\remarkname}{Remark}
\providecommand{\theoremname}{Theorem}

\begin{document}

\title{On the Fundamental Relationship Determining the Capacity of Static
and Mobile Wireless Networks}

\author{Guoqiang Mao, \emph{Senior Member, IEEE }%
\thanks{G. Mao is with the School of Electrical and Information Engineering,
the University of Sydney and National ICT Australia. %
}\emph{}%
\thanks{This research is supported by ARC Discovery projects DP110100538 and
DP120102030.%
}}
\maketitle
\begin{abstract}
Studying the capacity of wireless multi-hop networks is an important
problem and extensive research has been done in the area. In this
letter, we sift through various capacity-impacting parameters and
show that the capacity of both static and mobile networks is fundamentally
determined by the average number of simultaneous transmissions, the
link capacity and the average number of transmissions required to
deliver a packet to its destination. We then use this result to explain
and help to better understand existing results on the capacities of
static networks, mobile networks and hybrid networks and the multicast
capacity.\end{abstract}
\begin{IEEEkeywords}
Capacity, mobile networks, wireless networks
\end{IEEEkeywords}

\section{Introduction\label{sec:Introduction}}

Wireless multi-hop networks, in various forms, e.g. wireless sensor
networks, underwater networks, vehicular networks, mesh networks and
unmanned aerial vehicle formations, and under various names, e.g.
ad-hoc networks, hybrid networks, delay tolerant networks and intermittently
connected networks, are being increasingly used in military and civilian
applications. 

Studying the capacity of these networks is an important problem. Since
the seminal work of Gupta and Kumar \cite{Gupta00the}, extensive
research has been done in the area. Particularly, it was shown in
\cite{Gupta00the} that in an ad-hoc network with a total of $n$
nodes uniformly and \emph{i.i.d.} on an area of unit size and each
node is capable of transmitting at $W$ bits/s and using a fixed and
identical transmission range, the achievable per-node throughput,
when each node randomly and independently chooses another node in
the network as its destination, is given by $\lambda\left(n\right)=\Theta\left(\frac{W}{\sqrt{n\log n}}\right)$.
When the nodes are optimally and deterministically placed to maximize
throughput, the achievable per-node throughput becomes $\lambda\left(n\right)=\Theta\left(\frac{W}{\sqrt{n}}\right)$.
In \cite{Franceschetti07Closing}, Franceschetti \emph{et al.} considered
the same random network as that in \cite{Gupta00the} except that
nodes in the network are allowed to use two different transmission
ranges. They showed that by having each source-destination pair transmitting
using the ``highway system'', formed by nodes using the smaller
transmission range, the per-node throughput can also reach $\lambda\left(n\right)=\Theta\left(\frac{1}{\sqrt{n}}\right)$
even when nodes are randomly deployed. The existence of such highway
was established using the percolation theory \cite{Meester96Continuum}.
In \cite{Grossglauser02Mobility} Grossglauser and Tse showed that
in mobile networks, by leveraging on the nodes' mobility, a per-node
throughput of $\Theta\left(1\right)$ can be achieved at the expense
of unbounded delay. Their work \cite{Grossglauser02Mobility} has
sparked huge interest in studying the capacity-delay tradeoffs in
mobile networks assuming various mobility models and the obtained
results often vary greatly with the mobility models being considered,
see \cite{Neely05Capacity} for an example. Further, there is also
a significant amount of work studying the impact of infrastructure
nodes \cite{Zemlianov05Capacity} and multiple-access protocol \cite{Chau11Capacity}
on capacity and the multicast capacity \cite{Li09Multicast}. We refer
readers to \cite{Haenggi09Stochastic} for a more comprehensive review
of related work.

In this letter, we sift through these capacity-impacting parameters,
e.g. routing protocols, traffic distribution, mobility, presence of
infrastructure nodes, multiple-access protocol and scheduling algorithm,
and find the fundamental relationship determining the capacity of
both static and mobile networks. Specifically, considering a very
generic network setting, we show that the network capacity is fundamentally
determined by the link capacity, the average number of simultaneous
transmissions, and the average number of transmissions required to
deliver a packet to its destination. We then show how to use the result
to explain and better understand existing capacity results \cite{Gupta00the,Franceschetti07Closing,Grossglauser02Mobility,Neely05Capacity,Zemlianov05Capacity,Li09Multicast,Chau11Capacity}.

\section{Capacity of Static and Mobile Networks\label{sec:Capacity-of-Static-Networks}}

In this section, we establish the main result on the network capacity.
Specifically, consider a total of $n$ nodes distributed in a bounded
area $A$. These nodes may be either mobile or stationary. Packets
are transmitted between a source and its destination via multiple
intermediate relay nodes. Each node can be either a source, a relay,
a destination or a mixture. Let $V$ be the node set. Let $v_{i}\in V$
be a source node and let $b_{i,j}$ be the $j^{th}$ bit transmitted
from $v_{i}$ to its destination. Let $d\left(v_{i},j\right)$ be
the destination of $b_{i,j}$. For unicast transmission, $d\left(v_{i},j\right)$
represents a single destination; for multicast transmission, $d\left(v_{i},j\right)$
represents the set of all destinations of $b_{i,j}$. Let $h_{i,j}$
be the number of transmissions required to deliver $b_{i,j}$ to its
destination (or all destination nodes in $d\left(v_{i},j\right)$).
Let $\tau_{i,j,l}$, $1\leq l\leq h_{i,j}$ be the time required to
transmit $b_{i,j}$ in the $l^{th}$ transmission and assume that
the transmitting node is active during the entire $\tau_{i,j,l}$
interval. Let $Y_{t}$ be the number of simultaneous transmissions
in the network at time $t$. Let $N_{i,T}$ be the number of bits
transmitted by $v_{i}$ \emph{and} which reached their respective
destination during a time interval $\left[0,T\right]$, with $T$
being a large but arbitrary number. The network capacity, denoted
by $\eta\left(n\right)$, is defined as:

\begin{equation}
\eta\left(n\right)\triangleq\lim_{T\rightarrow\infty}\frac{\sum_{i=1}^{n}N_{i,T}}{T}\label{eq:definition of capacity}
\end{equation}

Note that the routing protocol used in the network plays an important
role in determining $\eta\left(n\right)$ and other parameters like
$h_{i,j}$ and $Y_{t}$. The validity of analytical results established
in this section however does not depend on the particular routing
protocol being used. Therefore we do not assume the use of a particular
routing protocol in the network. 

The average number of transmissions required to deliver a randomly
chosen bit to its destination, denoted by $k\left(n\right)$, equals
\begin{eqnarray}
k\left(n\right) & = & \lim_{T\rightarrow\infty}\frac{\sum_{i=1}^{n}\sum_{j=1}^{N_{i,T}}h_{i,j}}{\sum_{i=1}^{n}N_{i,T}}\label{eq:definition of average number of transmissions}
\end{eqnarray}
When $T$ is sufficiently large and the network is \emph{stable},
the amount of traffic in transit is negligible compared with the amount
of traffic that has already reached its destination. Therefore, the
following relationship can be established:

\begin{equation}
\lim_{T\rightarrow\infty}\frac{\sum_{i=1}^{n}\sum_{j=1}^{N_{i,T}}\sum_{l=1}^{h_{i,j}}\tau_{i,j,l}}{\int_{0}^{T}Y_{t}dt}=1\label{eq:fundamental relation between capacity and average transmission}
\end{equation}
A network is called stable if for any fixed $n$, assuming that each
node has an infinite queue, the queue length in any intermediate relay
node storing packets in transit does not grow towards infinity as
$T\rightarrow\infty$. 

Assuming that each node transmits at a fixed capacity $W$, then $\tau_{i,j,l}=\frac{1}{W}$.
It can be shown that
\begin{equation}
\lim_{T\rightarrow\infty}\sum_{i=1}^{n}\sum_{j=1}^{N_{i,T}}\sum_{l=1}^{h_{i,j}}\tau_{i,j,l}=\frac{1}{W}\lim_{T\rightarrow\infty}\sum_{i=1}^{n}\sum_{j=1}^{N_{i,T}}h_{i,j}\label{eq:average transmissions}
\end{equation}
Further, let 
\begin{equation}
E\left(Y\right)\triangleq\lim_{T\rightarrow\infty}\frac{\int_{0}^{T}Y_{t}dt}{T}\label{eq:definition of EY}
\end{equation}
where $E\left(Y\right)$ has the meaning of being the average number
of simultaneous transmissions in the network. It then follows from
(\ref{eq:definition of capacity}), (\ref{eq:definition of average number of transmissions}),
(\ref{eq:fundamental relation between capacity and average transmission}),
(\ref{eq:average transmissions}) and (\ref{eq:definition of EY})
that 
\begin{equation}
\eta\left(n\right)=\frac{E\left(Y\right)W}{k\left(n\right)}\label{eq:Capacity of mobile and static networks}
\end{equation}

\begin{rem}
The techniques used in obtaining Equation (\ref{eq:fundamental relation between capacity and average transmission})
and subsequently Equation (\ref{eq:Capacity of mobile and static networks})
is based on first considering transmissions in the network on the
individual node level by aggregating the transmissions at different
nodes, i.e. $\sum_{i=1}^{n}\sum_{j=1}^{N_{i,T}}\sum_{l=1}^{h_{i,j}}\tau_{i,j,l}$
and then evaluating transmissions in the network on the network level
by considering the number of simultaneous transmissions in the entire
network, viz. $\int_{0}^{T}Y_{t}dt$. Equation (\ref{eq:Capacity of mobile and static networks})
can also be obtained using Little's formula in queueing theory \cite{Kleinrock75Queueing}.
\end{rem}
Equation (\ref{eq:Capacity of mobile and static networks}) is obtained
under a very generic setting and is applicable for network of any
size. It reveals that the network capacity is fundamentally determined
by the average number of simultaneous transmissions $E\left(Y\right)$,
the average number of transmissions required for reaching the destination
$k\left(n\right)$ and the link capacity $W$. The two parameters
$E\left(Y\right)$ and $k\left(n\right)$ are often related. For example,
in a network where each node transmits using a fixed transmission
range $r\left(n\right)$, reducing $r\left(n\right)$ (while keeping
the network connected) will cause increases in both $E\left(Y\right)$
and $k\left(n\right)$ and the converse. On the other hand, $E\left(Y\right)$
and $k\left(n\right)$ also have their independent significance, and
can be optimized and studied independently of each other. For example,
an optimally designed routing algorithm can distribute traffic evenly
and avoid creating bottlenecks which helps to significantly increase
$E\left(Y\right)$ at the expense of slightly increased $k\left(n\right)$,
compared with shortest-path routing. Further, observing that each
transmission will ``consume'' a disk area of radius at least $\frac{Cr\left(n\right)}{2}$
in the sense that two simultaneous active transmitters must be separated
by an Euclidean distance of at least $Cr\left(n\right)$, where $C>1$
is a constant determined by the interference model \cite{Gupta00the},
the problem of finding the maximum number of simultaneous transmissions,
viz. an upper bound on $E\left(Y\right)$, can be converted into one
that finds the maximum number of non-overlapping equal-radius circles
that can be packed into $A$ and then studied as a densest circle
packing problem (see \cite{Yang12Connectivity} for an example). $E\left(Y\right)$
can also be studied as the transmission capacity of networks \cite{Weber10An}.
For unicast transmission, $k\left(n\right)$ becomes the average number
of hops between two randomly chosen source-destination pairs and has
been studied extensively \cite{Mao10Probability}. As will also be
shown in Section \ref{sec:Applicability of the result}, $E\left(Y\right)$
and $k\left(n\right)$ can be optimized separately to maximize the
network capacity.

We also note an important special case of (\ref{eq:Capacity of mobile and static networks}):
when the total number of source-destination pairs equals to $m$ and
each source-destination pair equally shares the network capacity,
the throughput per source-destination pair, denoted by $\lambda\left(n\right)$,
is given by 
\begin{equation}
\lambda\left(n\right)=\frac{E\left(Y\right)W}{mk\left(n\right)}\label{eq: per source-destination throughput}
\end{equation}
The total number of possible source-destination pairs in the network
equals to $n\left(n-1\right)$ and if each node randomly and independently
chooses another node in the network as its destination, as considered
in \cite{Gupta00the,Franceschetti07Closing,Grossglauser02Mobility,Neely05Capacity,Zemlianov05Capacity,Chau11Capacity},
$m=n$.

\section{Applications of (\ref{eq:Capacity of mobile and static networks})
to Existing Results\label{sec:Applicability of the result}}

In this section, we use the result on network capacity established
in (\ref{eq:Capacity of mobile and static networks}) and (\ref{eq: per source-destination throughput})
to explain and better understand existing results \cite{Gupta00the,Franceschetti07Closing,Grossglauser02Mobility,Neely05Capacity,Zemlianov05Capacity,Li09Multicast,Chau11Capacity}
in the area. Unless otherwise specified, we consider a network with
$n$ nodes uniformly and \emph{i.i.d.} on a unit square $A$ and each
node is capable of transmitting at a fixed rate of $W$ bits/s. A
node chooses its destination randomly and independently of other nodes
and the total number of source-destination pairs equals to $n$, viz.
$m=n$. In some literature \cite{Franceschetti07Closing,Zemlianov05Capacity,Li09Multicast},
a different network area is considered and their results are converted
into a unit square and discussed.

\subsection{Static ad-hoc networks}

In \cite{Gupta00the}, Gupta and Kumar first considered the network
defined above and that each node transmits using a fixed and identical
transmission range $r\left(n\right)$. Given the above setting, it
is straightforward to establish that $E\left(Y\right)=\Theta\left(\frac{1}{r^{2}\left(n\right)}\right)$
(as pointed out in Section \ref{sec:Capacity-of-Static-Networks},
each transmission consumes a disk area of radius $\Theta\left(r\left(n\right)\right)$)
and $k\left(n\right)=\Theta\left(\frac{1}{r\left(n\right)}\right)$.
Using (\ref{eq: per source-destination throughput}) and noting that
$m=n$, it can be shown that $\lambda\left(n\right)=\Theta\left(\frac{W}{nr\left(n\right)}\right)$,
viz. a smaller transmission range will result in a larger throughput.
The minimum transmission range required for the network to be connected
is well known to be $r\left(n\right)=\Theta\left(\sqrt{\frac{\log n}{n}}\right)$.
Accordingly, the per-node throughput becomes $\lambda\left(n\right)=\Theta\left(\frac{W}{\sqrt{n\log n}}\right)$.
By placing nodes optimally (e.g. on grid points) however, the transmission
range required for a connected network reduces to $r\left(n\right)=\Theta\left(\frac{1}{\sqrt{n}}\right)$.
Thus the per-node throughput becomes $\lambda\left(n\right)=\Theta\left(\frac{W}{\sqrt{n}}\right).$
Therefore the $\frac{1}{\sqrt{\log n}}$ factor is the price in reduction
of network capacity to pay for placing nodes randomly, instead of
optimally.

In the networks considered by Franceschetti \emph{et al. }\cite{Franceschetti07Closing},
two transmission ranges are allowed, viz. a smaller transmission range
of $\Theta\left(\frac{1}{\sqrt{n}}\right)$ for nodes forming the
highway and a larger transmission range of $r\left(n\right)=\Theta\left(\sqrt{\frac{\log n}{n}}\right)$
for ordinary nodes. Most transmissions are through the highway using
the smaller transmission range while the larger transmission range
is only used for the last mile, i.e. between a source (or destination)
and its nearest highway node. Therefore both $E\left(Y\right)$ and
$k\left(n\right)$ are dominated by the smaller transmission range
and accordingly $E\left(Y\right)=\Theta\left(n\right)$ and $k\left(n\right)=\Theta\left(\sqrt{n}\right)$.
It then readily follows that $\lambda\left(n\right)=\Theta\left(\frac{1}{\sqrt{n}}\right)$. 

Observing that in a large network, a much smaller transmission range
is required to connect most nodes in the network (i.e. forming a giant
component) whereas the larger transmission range of $\Theta\left(\sqrt{\frac{\log n}{n}}\right)$
is only required to connect the few hard-to-reach nodes \cite{Ta09On},
a routing scheme can be designed, which achieves a per-node throughput
of $\lambda\left(n\right)=\Theta\left(\frac{1}{\sqrt{n}}\right)$
and does not have to use the highway system, such that a node uses
smaller transmission ranges for most communications and only uses
a larger transmission if the next-hop node cannot be reached when
using smaller transmission ranges.

\subsection{Mobile ad-hoc networks}

In the mobile ad-hoc networks considered in \cite{Grossglauser02Mobility},
nodes are mobile and the spatial distribution of nodes is stationary
and ergodic with stationary distribution uniform on $A$. Moreover,
the trajectories of different nodes are i.i.d. A two-hop relaying
strategy is adopted. In the first stage, a source transmits a packet
to its nearest neighbor (acting as a relay). As the source moves around,
different packets are transmitted to different relay nodes. In the
second stage, either the source or a relay transmits the packet to
the destination when it is closest to the destination. 

Obviously the two-hop relaying strategy helps to cap $k\left(n\right)$
at $2$. Compared with a one-hop strategy where a source is only allowed
to transmit when it is close to its destination, the two-hop relaying
strategy also helps to spread the traffic stream between a source-destination
pair to a large number of intermediate relay nodes such that in steady
state, the packets of every source node will be distributed across
all the nodes in the network. This arrangement ensures that every
node in the network will have packets buffered for every other node.
Therefore a node always has a packet to send when a transmission opportunity
is available. In this way, $E\left(Y\right)$ is also maximized. Since
the Euclidean distance between a node and its nearest neighbor is
$\Theta\left(\frac{1}{\sqrt{n}}\right)$, it follows that $E\left(Y\right)=\Theta\left(n\right)$
\cite{Grossglauser02Mobility}. As an easy consequence of (\ref{eq:Capacity of mobile and static networks})
and (\ref{eq: per source-destination throughput}), $\lambda\left(n\right)=\Theta\left(1\right)$
and $\eta\left(n\right)=\Theta\left(n\right)$. Capacity of mobile
ad-hoc networks assuming other mobility models and routing strategies
\cite{Neely05Capacity} can also be obtained analogously.

Given the insight revealed in (\ref{eq:Capacity of mobile and static networks})
and (\ref{eq: per source-destination throughput}), it can be readily
shown that in a network with a different traffic model than that in
\cite{Grossglauser02Mobility}, viz. each node has an infinite stream
of packets for every other node in the network, a one-hop strategy
can also achieve a network capacity of $\eta\left(n\right)=\Theta\left(n\right)$.
Therefore the insight revealed in (\ref{eq:Capacity of mobile and static networks})
and (\ref{eq: per source-destination throughput}) helps to design
the optimum routing strategy for different scenarios of mobile ad-hoc
networks.

\subsection{Multicast capacity}

Now we consider the multicast capacity of a network using a similar
setting as that in \cite{Li09Multicast}. It is assumed that all nodes
use the same transmission range $r\left(n\right)=\Theta\left(\sqrt{\frac{\log n}{n}}\right)$.
Each node chooses a set of $l-1$ points randomly and independently
from $A$ and multicast its data to the nearest node of each point.
Further, it is assumed that the multicast transmission from each source
follows the path of the Euclidean minimum spanning tree rooted at
the source. Let $\vartheta_{i}$ be the rate at which $v_{i}$ send
data to its destination nodes. The multicast capacity of the network
is defined as $\eta\left(n\right)=\sum_{v_{i}\in V}\vartheta_{i}$,
which is consistent with the definition in (\ref{eq:definition of capacity}).

According to the analysis in \cite{Li09Multicast}, when $l=O\left(\frac{n}{\log n}\right)$,
the number of transmissions required to reach all $l-1$ multicast
destinations is $k\left(n\right)=\Theta\left(\frac{\sqrt{l}}{r\left(n\right)}\right)$.
$E\left(Y\right)$ is mainly determined by the transmission range
$r\left(n\right)$ and is little affected by the change to multicast.
Therefore,
\[
\eta\left(n\right)=\Theta\left(\frac{1}{r^{2}\left(n\right)}\times\frac{r\left(n\right)}{\sqrt{l}}W\right)=\Theta\left(W\sqrt{\frac{\log n}{ln}}\right)
\]
When $l=\Omega\left(\frac{n}{\log n}\right)$, the density of the
multicast destination nodes becomes high enough such that the probability
that a single transmission will deliver the data to more than one
destination nodes becomes high. Consequently $k\left(n\right)=\Theta\left(\frac{1}{r^{2}\left(n\right)}\right)$
(i.e. the number of transmissions required to cover the entire network)
and $\eta\left(n\right)=\Theta\left(W\right)$.

\subsection{Hybrid networks}

Now we consider the impact of infrastructure nodes. In addition to
$n$ ordinary nodes, a set of $M$ infrastructure nodes are regularly
or randomly placed on $A$ where $M<n$. These infrastructure nodes
act as relay nodes only and do not generate their own traffic. Following
a similar setting as that in \cite{Zemlianov05Capacity}, it is assumed
that the infrastructure nodes have the same transmission range $r\left(n\right)=\Theta\left(\sqrt{\frac{\log n}{n}}\right)$
and bandwidth $W$ when they communicate with the ordinary nodes and
these infrastructure nodes are inter-connected via a backbone network
with much higher bandwidth. Further, it is assumed that the routing
algorithm has been optimized such that these infrastructure nodes
do not become the bottleneck, which may be possibly caused by a poorly
designed routing algorithm diverting excessive amount of traffic to
the infrastructure nodes.

First consider the case that $M=o\left(\frac{1}{r^{2}\left(n\right)}\right)=o\left(\frac{n}{\log n}\right)$.
In this situation, the number of transmissions involving an infrastructure
node as a transmitter or receiver is small and has little impact on
$E\left(Y\right)$. Further, it can be shown that the average Euclidean
distance between a randomly chosen pair of infrastructure nodes is
$\Theta\left(1\right)$ \cite{Philip07The}. That is, a packet transmitted
between two infrastructure nodes moves by an Euclidean distance of
$\Theta\left(1\right)$ whereas a packet transmitted by a pair of
directly connected ordinary nodes moves by an Euclidean distance of
$\Theta\left(r\left(n\right)\right)$. Therefore a transmission between
two infrastructure nodes \emph{is equivalent to} $\Theta\left(\frac{1}{r\left(n\right)}\right)$
transmissions between ordinary nodes and the \emph{equivalent }average
number of simultaneous ordinary node transmissions equals to $\Theta\left(\left(\frac{1}{r^{2}\left(n\right)}-M\right)+\frac{M}{r\left(n\right)}\right)=\Theta\left(\frac{1}{r^{2}\left(n\right)}+\frac{M}{r\left(n\right)}\right)$.
It follows that
\[
\eta\left(n\right)=\Theta\left(\frac{\left(\frac{1}{r^{2}\left(n\right)}+\frac{M}{r\left(n\right)}\right)W}{\frac{1}{r\left(n\right)}}\right)=\Theta\left(\left(\sqrt{\frac{n}{\log n}}+M\right)W\right)
\]
Therefore when $M=o\left(\sqrt{\frac{n}{\log n}}\right)$, the infrastructure
nodes have little impact on the order of $\eta\left(n\right)$; when
$M=\Omega\left(\sqrt{\frac{n}{\log n}}\right)$, the infrastructure
nodes start to have dominant impact on the network capacity and the
above equation reduces to $\eta\left(n\right)=\Theta\left(MW\right)$.
Noting that the fundamental reason why infrastructure nodes improve
capacity is that they help a pair of ordinary nodes separated by a
large Euclidean distance to leapfrog some very long hops. Therefore
the same result in the above equation can also be obtained by analyzing
the reduction in $k\left(n\right)$ directly. The analysis is albeit
more complicated.

When $M=\Omega\left(\frac{n}{\log n}\right)$, the number of simultaneous
active infrastructure nodes becomes limited by the transmission range.
More specifically, only $\Theta\left(\frac{1}{r^{2}\left(n\right)}\right)=\Theta\left(\frac{n}{\log n}\right)$
infrastructure nodes can be active simultaneously. Further, each ordinary
node can access its nearest infrastructure node in $\Theta\left(1\right)$
hops. Therefore $\eta\left(n\right)=\Theta\left(\frac{nW}{\log n}\right)$.

The above results are consistent with the results in \cite{Zemlianov05Capacity}. 

However we further note that when $M=\Omega\left(\frac{n}{\log n}\right),$
a smaller transmission range of $\Theta\left(\frac{1}{\sqrt{M}}\right)$
is sufficient for an ordinary node to reach its nearest infrastructure
node and hence achieve connectivity. A smaller transmission range
helps to maximize $E\left(Y\right)$ while $k\left(n\right)=\Theta\left(1\right)$.
Therefore the achievable network capacity using the smaller transmission
range is $\eta\left(n\right)=\Theta\left(MW\right)=\Omega\left(\frac{nW}{\log n}\right)$,
which is better than the result $\eta\left(n\right)=\Theta\left(\frac{nW}{\log n}\right)$
in \cite{Zemlianov05Capacity}. Moreover, different from the conclusion
in \cite{Zemlianov05Capacity} suggesting that when $M=\Omega\left(\frac{n}{\log n}\right)$,
further investment in infrastructure nodes will not lead to improvement
in capacity, our result suggests that even when $M=\Omega\left(\frac{n}{\log n}\right),$
capacity still keeps increasing linearly with $M$. This capacity
improvement is achieved by reducing the transmission range with the
increase in $M$.

\bibliographystyle{ieeetr}

\end{document}